\documentclass[prc,nofootinbib,preprintnumbers]{revtex4}[12pt]
\usepackage{graphicx}
\usepackage{bm}
\usepackage{amssymb,amsmath}
\usepackage{epstopdf}
\usepackage{epsfig,dcolumn}
\begin{document}

\preprint{MITP/13-074}
\title{Theory Uncertainty in Extracting the Proton's Weak Charge: White Paper}

\author{Mikhail Gorchtein} 
\email{gorshtey@kph.uni-mainz.de} 
\affiliation{PRISMA Cluster of Excellence, Institut f\"ur Kernphysik, Universit\"at Mainz, D-55128 Mainz, Germany} 

\author{Jens Erler}
\email{erler@fisica.unam.mx}
\affiliation{Departamento de F\'isica Te\'orica, Instituto de F\'isica, 
Universidad Nacional Aut\'onoma de M\'exico, M\'exico D.F. 04510, M\'exico}

\author{Tobias Hurth}
\email{hurth@uni-mainz.de}
\author{Hubert Spiesberger}
\email{spiesber@uni-mainz.de}
\affiliation{PRISMA Cluster of Excellence, Institut f\"ur Physik, Universit\"at Mainz, D-55099 Mainz, Germany}

\author{Krishna Kumar}
\email{kkumar@physics.umass.edu}
\author{Michael J.~Ramsey-Musolf}
\email{mjrm@physics.umass.edu}
\affiliation{Amherst Center for Fundamental Interactions, Department of Physics,
University of Massachusetts Amherst, Amherst, MA 01003, USA}

\author{Harvey B.~Meyer}
\email{meyerh@kph.uni-mainz.de}
\affiliation{PRISMA Cluster of Excellence, Institut f\"ur Kernphysik and Helmholtz Institut Mainz, 
Universit\"at Mainz, D-55099 Mainz, Germany}

\begin{abstract}
{We review the state-of-the-art and address open questions relative to the theory uncertainty 
of the $\gamma-Z$ box contribution to the $A_{PV}$ measurement within the QWEAK experiment at 
Jefferson Lab. This white paper summarizes the contributions by participants
and discussion sessions on this topic within the 
MITP Workshop on Precision Electroweak Physics held in 
Mainz, Germany, September 23 - October 11, 2013
\url{http://indico.cern.ch/conferenceDisplay.py?confId=248072}. }
\end{abstract}

\maketitle

The running of the weak mixing angle \cite{erler1,erler2} is a
distinct prediction of the Standard Model. This running is the subject
of an extensive experimental program in atomic, nuclear, neutrino and
collider physics. 
The Qweak experiment at Jefferson Lab \cite{qweak} aims at a 0.3\%
determination of $\sin^2\theta_W(0)$. Translated into {\it e. g.} masses of 
hypothetical heavy New Physics (NP) particles,
this measurement will probe NP contributions from scales beyond 1 TeV. 
It is achieved by means of a 2.5\%
measurement of the parity violating asymmetry ($A_{PV}$) in elastic $e$-$p$
scattering of $1.165$ GeV polarized electrons off a hydrogen
target at a scattering angle of $8^\circ$, which combined with the theory uncertainty will yield a 4\% 
determination of the proton's weak charge~\cite{marciano,mjrm,young}. 

Recently, the calculation of the $\gamma Z$-box contribution 
was re-scrutinized by means of forward dispersion relations
\cite{gorchtein1,sibirtsev,carlson1,hall,carlson2,gorchtein2}, and it
was found that a 7\% effect (relative to the SM-predicted value of the
 $A_{PV}$ in the Qweak kinematics) 
 was missed in the original analysis. The three groups
\cite{gorchtein2,carlson2,hall} agree on the size of the correction,
however the uncertainty estimates differ: Hall {\it et~al.\/} claim
Re$\Box_{\gamma Z}^V=(5.57\pm0.36)\times10^{-3}$, Carlson and Rislow claim
Re$\Box_{\gamma Z}^V=(5.7\pm0.9)\times10^{-3}$, and Gorchtein {\it et~al.\/} claim
Re$\Box_{\gamma Z}^V=(5.4\pm2.0)\times10^{-3}$, thus the error
estimates may differ by a factor of 6. The largest uncertainty 
estimate \cite{gorchtein2} constitutes 2.6\% of the SM value of the
$A_{PV}$, thus clearly interfering with the overall
2\% theory uncertainty \cite{mjrm}. It is highly desirable to
provide a unified theory uncertainty on this correction before the
final analysis of the Qweak experiment is completed 
(Ref.~\cite{qweak} contains 4\% of data taken). 

The three groups agree that the most relevant kinematics within the
loop corresponds to $Q^2\leq3$ GeV$^2$ and $W\leq5$ GeV, and the
emphasis is on the lower part of these regions. 
The electromagnetic data for this region in the $Q^2$-$W$ plane feature
resonances and the non-resonant background. 
All three models share one common property, 
the parametrization of the resonance data by Christy and Bosted
\cite{christybosted} in terms of 7 resonances and a smooth
background. 
As a first step all groups perform an isospin decomposition of
the resonance part of the inclusive virtual photo-absorption data. This
procedure is model dependent since i) the data may be fitted with a
different number of resonances, and ii) the resonances may be
identified differently based on the inclusive data alone.
In a second step, the resonances' strength in the $\gamma Z$
interference cross section are predicted from their strength in the
electromagnetic cross section. This is done based on the resonances'
quantum numbers obtained with the identification in the first step,
plus the conservation of the vector current (CVC). 
These two steps are largely in agreement between all three calculations.

For the background, experimental data cannot be directly decomposed into definite isospin channels. 
As a result, any such decomposition is model-dependent. 
For (leading twist) DIS that is generally warranted at
substantially high $Q^2$ and large Bjorken-$x$, the $\gamma(Z)$ couples perturbatively
to quarks within the proton and the isospin/flavor structure is obtained
from tree-level quark charges. This is the picture that Hall et al \cite{hall}
chose to constrain the isospin structure of the background. 
At high energy and low
virtualities good qualitative understanding of the data 
is offered by the Vector Dominance Model (VDM) where a
photon fluctuates into vector meson (VM) states
($\rho^0,\omega,\phi,\dots$) that then interact with the proton. Na\"ive 
VDM assumes that the photon wave function can be decomposed into a
basis of vector meson states, and that this basis is complete and
orthogonal. These assumptions lead to the VDM sum rule that expresses the total
photo-absorption cross section at a given energy through forward
differential cross sections for VM photo-production. 
This sum rule was measured at DESY with 82 GeV
photons, and revealed a 21\% incompleteness of the VM basis. The
missing piece remains unidentified and is called ``continuum".
VDM also extends this decomposition to virtual photons: each
flavor contribution has natural $Q^2$-dependence due to the respective
VM propagator, thus leading to dipole form in the cross section, 
$\sim1/(1+Q^2/m_V^2)^2$. The continuum part is supplemented with a
phenomenological form factor chosen such as to fit virtual
photo-absorption data. For instance, for the transverse cross section it
requires roughly a monopole form factor $\sim1/(1+Q^2/m_0^2)$ with
$m_0\sim1.5$ GeV, the approach that leads to reasonable description of
data below $Q^2=1.5-2$ GeV$^2$ (see Ref \cite{alwall}). 
The VDM-based flavor decomposition
of the inclusive cross section deteriorates when going to virtual
photons: the sum of VM pieces becomes of the same size as the continuum
at $Q^2\sim0.8$ GeV$^2$.

At the next step, to obtain the $\gamma-Z$ interference cross section
we perform the isospin rotation of the background according to the
isospin/flavor decomposition performed in the previous step. The
presence of the continuum contribution of an unidentified flavor
content sets the limit on the precision of such a procedure. It is the
different treatment of this uncertainty by the three groups that leads
to differences in the $\gamma Z$-box uncertainty estimate for the QWEAK. 

Gorchtein et al \cite{gorchtein2} assign a conservative 100\%
uncertainty to the continuum contribution, resulting in 
35\% uncertainty relative to the size of $\Box_{\gamma Z}^V$. 
Hall et al. \cite{hall} proposed to match the uncertainty due to the
continuum to that of the DIS data at $Q^2\geq2.5$ GeV$^2$. Given the
quality of those data and assuming this uncertainty to stay constant
all the way down to $Q^2=0$ one obtains a 6\% uncertainty relative to the size of $\Box_{\gamma Z}^V$
\cite{hall}. In view of the 21\% uncertainty at $Q^2=0$ that is built in the VDM, 
all sources of uncertainty entering this treatment should be carefully assessed. 
These two approaches give the upper and the lower limits for the
uncertainty of $\Box_{\gamma Z}^V$. 

Ref.~\cite{carlson2} quotes an
uncertainty that lies between the two extremes. It follows from the 
$SU(4)$ and $SU(6)$ versions of the non-relativistic constituent quark model (NRCQM), 
and derives the uncertainty from the difference between these two
cases. The NRCQM is not expected to be precise at very low $Q^2$ as it
misses the pion degrees of freedom, although at some intermediate
$Q^2$ it may give reasonable description of the hadronic excitation spectrum. It
is difficult to assess whether all systematic uncertainties of the NRCQM are reflected in 
the estimate of Ref.~\cite{carlson2}.

In the context of the upcoming experimental programs at Jefferson Lab and Mainz, 
we agreed that
\begin{itemize}
\item{} the potential impact of auxiliary measurements of $A_{PV}$ over relevant ranges of low $Q^2$ and moderate W in the future PV program at JLab (6 GeV PVDIS \cite{pvdis}, SOLID \cite{solid} and MOLLER \cite{moller}) on the uncertainty of $\Box_{\gamma Z}$ needs to be reviewed and evaluated;\\
\item{} a measurement of $Q_W^p$ with an electron energy below 200 MeV 
is planned at Mainz MESA/P2 to minimize the uncertainty due to the 
  $\Box_{\gamma Z}^V$ correction. It aims at 1\% measurement of the PV asymmetry,
  so that constraining the theoretical uncertainty to below 1\% is
  mandatory. 
\end{itemize}

In view of an upcoming topical meeting at JLab dedicated to
constraining the theory uncertainty on the $\Box_{\gamma Z}^V$
calculation we propose our vision on the questions that need be
addressed in order to come to a realistic theory error on this correction.
\begin{itemize}
\item{} All three models use the same resonance data
  parametrization by Ref.~\cite{christybosted}. Can this lead to a common bias? 
A previous calculation by Sibirtsev et al. \cite{sibirtsev} has used a
different parametrization and obtained
$(4.7^{+1.1}_{-0.4})\times10^{-3}$. Although it agrees within the
errors with the three other evaluations it does not exclude such a
possibility.\\
\item{} The result of the isospin rotation of the electromagnetic
  resonance excitations relies on the identification of their quantum
  numbers plus uncertainties in their strength on the proton and the
  neutron. Is every step in this procedure reliable, and is there a
  way to further reduce this uncertainty?\\
\item{} The isospin rotation of the background relies on the
  isospin/flavor decomposition of the inclusive cross section
  according to the VDM. This decomposition was tested experimentally
  only at $Q^2=0$ and at $W=82$ GeV. Refs. \cite{gorchtein2,hall} use
  this experimental result, but such high values of $W$ are not
  representative of the $\Box_{\gamma Z}^V$ calculation in the QWEAK kinematics. 
  It is implicitly
  assumed that this flavor decomposition is a constant as function of
  $W$. This assumption has to be tested explicitly. Can available data
  on VM photo-production at JLab energies be utilized to test the VDM
  sum rule? If not, is it possible to propose a dedicated measurement
  at JLab? \\
\item{} The extrapolation of other PV measurements down to $Q^2=0$
  proposed in Ref.~\cite{young} (Fig. 2 of Ref.~\cite{qweak}) includes some 
  but not all effects of the $\Box_{\gamma Z}^V$. Are there
  possible issues with this extrapolation that are relevant for the
  $Q_W^p$ extraction?\\
\item{} Can the uncertainty in $\Box_{\gamma Z}^A$ be further reduced? 
\end{itemize}

{\it Summary}\\
We reviewed existing calculations of the $\Box^V_{\gamma Z}$ contribution to the 
parity-violating asymmetry in elastic electron-proton scattering in the kinematics of existing and 
upcoming experiments. All sources of theoretical uncertainties entering each of the evaluation 
were critically assessed. We put theoretical achievements and challenges in 
the context of current and upcoming experiments at Jefferson Lab and at Mainz to propose our vision of 
the medium-range to-do-list. Finally, on a shorter range, we propose a list of more urgent questions 
that should be addressed at the upcoming mini-workshop 
``$\gamma Z$ box(ing): Radiative corrections to parity-violating electron scattering" to be held December 16-17, 2013 at Jefferson Lab
\url{https://jlab.org/conferences/gz-box/}.

\acknowledgments{
We acknowledge all the contributions of the participants of the MITP workshop on Precision Electroweak Physics. 
M.G, H.S, H.B.M. and K.K. acknowledge support from the DFG within the SFB 1044.
J.E. was supported by PAPIIT (DGAPA--UNAM) project IN106913 and CONACyT (M\'exico) project 151234,
and gratefully acknowledges the hospitality and support by the Mainz Institute for Theoretical Physics (MITP) 
within the PRISMA Cluster of Excellence.}

\end{document}